\documentclass[aps,pra,twocolumn]{revtex4}   
\newcommand {\be}{\begin{equation}}
\newcommand {\ee}{\end{equation}}
\newcommand {\bey}{\begin{eqnarray}}
\newcommand {\eey}{\end{eqnarray}}
\usepackage{epsfig}
\begin{document}

\title{Thermalization of an impurity cloud in a Bose-Einstein condensate.}

\author{Alberto Montina}
\affiliation{Istituto Nazionale di Ottica Applicata, Largo E.~Fermi 6,50125 Firenze,
Italy}
\date{\today}

\begin{abstract}
We study the thermalization dynamics of an impurity cloud inside a Bose-Einstein 
condensate at finite temperature, introducing a suitable Boltzmann equation. Some
values of the temperature and of the initial impurity energy are considered.
We find that, below the Landau critical velocity, the macroscopic population of the
initial impurity state reduces its depletion rate because of stimulation effects. 
For sufficiently high velocities
the opposite effect occurs. For appropriate parameters the collisions cool the 
condensate. The maximum cooling per impurity atom is obtained with multiple 
collisions.
\end{abstract}

\maketitle
In a Bose-Einstein condensate (BEC) many atoms populate the same quantum state. This
system is, therefore, suitable to realize an atom laser~\cite{atomlaser}. That
is obtained
creating atoms in an untrapped or antitrapped level by means of a rf or Raman outcoupling.
Elastic collisions between impurity atoms and the parent BEC 
have been observed at MIT~\cite{ketterle}. These collisions can deplete~\cite{band} 
and heat the outcoupled atoms. 
In this work we study the
thermalization dynamics of the impurity atoms, created by a rf or Raman outcoupling. 
We find the following main results:
i) we evaluate the impurity thermalization and depletion time for some values of
the temperature and of the initial impurity energy. 
The thermalization is not suppressed below the Landau critical velocity~\cite{landau}
for a finite condensate temperature $T$, according to a previous work~\cite{montina};
ii) below the Landau critical velocity the macroscopic 
population of the atom laser state slows down the impurity depletion,
whereas, for sufficiently high velocities the macroscopic
population enhances the scattering rate, as observed in Ref.~\onlinecite{ketterle};
iii) the collisions cool the condensate if
the impurity initial energy $\epsilon_{in}$ is below its thermal equilibrium energy,
or even if this condition is not fulfilled, provided the thermal energy, divided by
the chemical potential, is sufficiently small.
These effects can be observed creating
a succession of atom laser packets and observing a reduction of the impurity 
depletion rate in the last packets. This series can be used
to reduce the condensate temperature, as proposed in Ref.~\onlinecite{montina}. 
In Ref.~\onlinecite{montina} we evaluated the scattering rate of an impurity
inside a condensate at finite temperature and found that for suitable initial 
velocities the impurity acquired energy, on average, after the first collision.
However processes with multiple collisions were not considered. Here we show that
the maximum cooling per impurity atom is obtained, for $T$ not too small, at 
thermal equilibrium, that is, with multiple collisions.
It is important to understand that the first cooling is obtained by a {\cal weak}
rf or Raman outcoupling. In fact, the condensed fraction for a trapped condensate
is $n_0/n_t=1-(T/T_0)^3$~\cite{dalfovo}, where $n_t$, $n_0$ are the
number of overall and condensed atoms, respectively, and $T_0$ is the transition
temperature. The number of non-condensed atoms is $n_t-n_0=\alpha T^3$, where
$\alpha\equiv n_t/T_0^3$ is a constant~\cite{dalfovo}. This means that
the condensate temperature is given by the number of non-condensed atoms, with no
regard to the number of the condensed ones. So, if the outcoupling does not 
create many impurities, the number of outcoupled non-condensed atoms is small and,
consequently, the impurity temperature is lower than the parent condensate one.
For this reason, it is possible to cool sympathetically a condensate creating a
small impurity cloud, as suggested in this work.

To simplify the problem we consider an homogeneous condensate, in fact it is meaningful
in a trapped system in the sense of the Thomas-Fermi approximation. 
If the number of outcoupled atoms is sufficiently low, the initial thermal energy of the 
impurities can be neglected. 
We first introduce a Boltzmann equation for an impurity atom. 
A similar equation in momentum space is obtained, for a single particle and $T=0$, in
Ref.~\onlinecite{timme}.
It is easy to show that the system goes toward the Boltzmann equilibrium distribution. 
We extend our discussion
to a cloud of impurity atoms introducing suitable bosonic stimulation terms. We obtain
an equation of the same type of the quantum Boltzmann equation~\cite{kagan,semikoz,holland}.
It is verified that the Bose-Einstein distribution is the steady state 
solution.

Let us consider an impurity atom in an uniform condensate at temperature $T$. 
If the impurity energy is $E_i$, the probability that the particle scatters in a 
state with energy $E_f$ is $\Gamma(E_f,E_i)=\Gamma_1(E_f,E_i)+\Gamma_2(E_f,E_i)$, 
where~\cite{montina}
\bey
&&\Gamma_{1,2}(E_f,E_i)=n_0\left(\frac{2\hbar a}{M}\right)^2\int dq d\Omega q^2 S(q)\times \\
\nonumber
&&\delta\left(\frac{E_i-E_f}{\hbar}\mp\omega_q^B\right) 
\delta\left(E_f-\frac{\hbar^2(\vec k_i\mp\vec q)^2}{2M}\right)
\frac{\pm1}{1-e^{\mp\beta\hbar\omega_b^B}},
\eey
$\hbar\vec k_i$, $1/\beta\equiv k_B T$, $S(q)\equiv \omega_q^0/\omega_q^B$, 
$\hbar\omega_q^0$ and
$\hbar\omega_q^B\equiv \sqrt{\hbar\omega_q^0(\hbar\omega_q^0+2\mu)}$ 
being, respectively, the initial impurity momentum, 
the condensate temperature in energy unit,
the static structure factor and the energies of a free particle and of a Bogoliubov 
quasiparticle
of momentum $q$. $M$, $\mu$ and $a$ are the atomic mass, the chemical potential and the
scattering length for s-wave collisions between the impurity atoms and the condensate ones.
$\mu$ is equal to $n_0 g$, where $n_0$ is the condensate density and $g=4\pi\hbar^2 a_0/M$,
$a_0$ being the condensate s-wave scattering length.

$\Gamma_1$ and $\Gamma_2$ are associated with dissipative and cooling processes, respectively.
In the first
case condensate phonons are created, in the second one they are annihilated~\cite{montina}.
With a little of algebra we find that 
$\Gamma(E_f,E_i)=2(\gamma/\mu) G(2 E_f/\mu,2 E_i/\mu)$, where 
$\gamma=4\pi n_0 a^2 c$, $c\equiv\sqrt{\mu/M}$ being the sound velocity, and
\bey
\nonumber
G(\epsilon_f,\epsilon_i)=\frac{1}{\sqrt{\epsilon_i}}\left[1-
\frac{1}{\sqrt{1+\Delta\epsilon^2/4}}\right]\times 
\frac{W(\epsilon_f,\epsilon_i)}{1-e^{\bar\beta(\epsilon_f-\epsilon_i)}},
\eey
with $\bar\beta=\beta\mu/2$ and $\Delta\epsilon=\epsilon_f-\epsilon_i$.
$W(\epsilon_f,\epsilon_i)$ is $1$ when $\epsilon_i\ge\epsilon_f\ge1/\epsilon_i$,
$-1$ when $\epsilon_f>\epsilon_i\ge1/\epsilon_f$
and zero elsewhere.
The Boltzmann equation is
\be\label{bolteq}
\frac{dP(\epsilon)}{d\tau}=\int G(\epsilon,\bar\epsilon)P(\bar\epsilon)d\bar\epsilon
-\int G(\bar\epsilon,\epsilon)P(\epsilon)d\bar\epsilon,
\ee
where $P(2E/\mu)$ is the density probability per unit energy and $\tau\equiv\gamma t$.
It is easy to demonstrate that $P(\epsilon)\propto \sqrt{\epsilon}e^{-\bar\beta\epsilon}$
is a steady solution of Eq.~(\ref{bolteq}). 

The system dynamics depends exclusively on the product $\beta\mu=\mu/(k_BT)$ and the 
initial probability distribution. The condensate density and the scattering length are 
present in the rate factor $\gamma$ and, therefore, they influence only the time 
scale of the dynamics.

\begin{figure}
\epsfig{figure=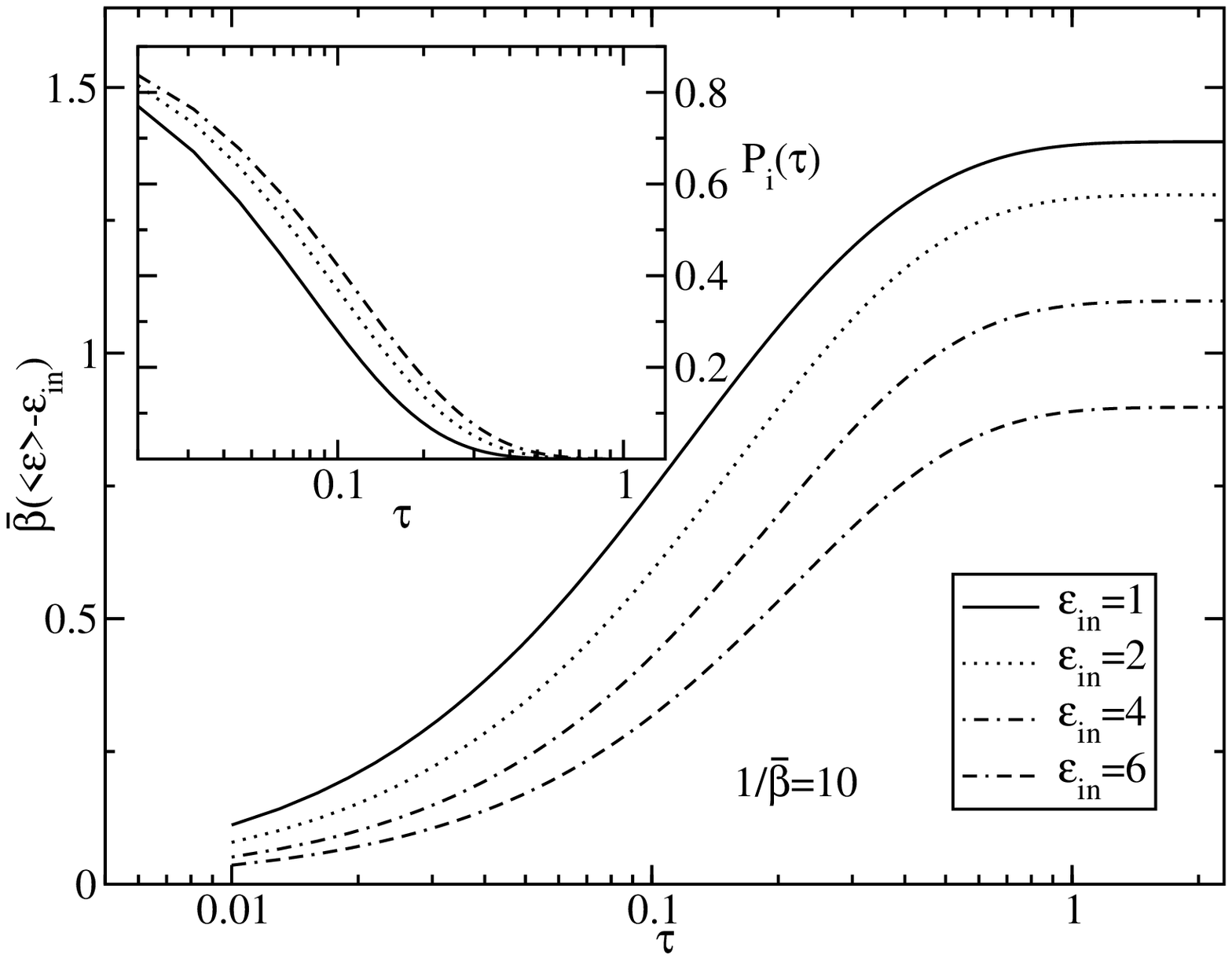,width=5.0cm,angle=-90}
\epsfig{figure=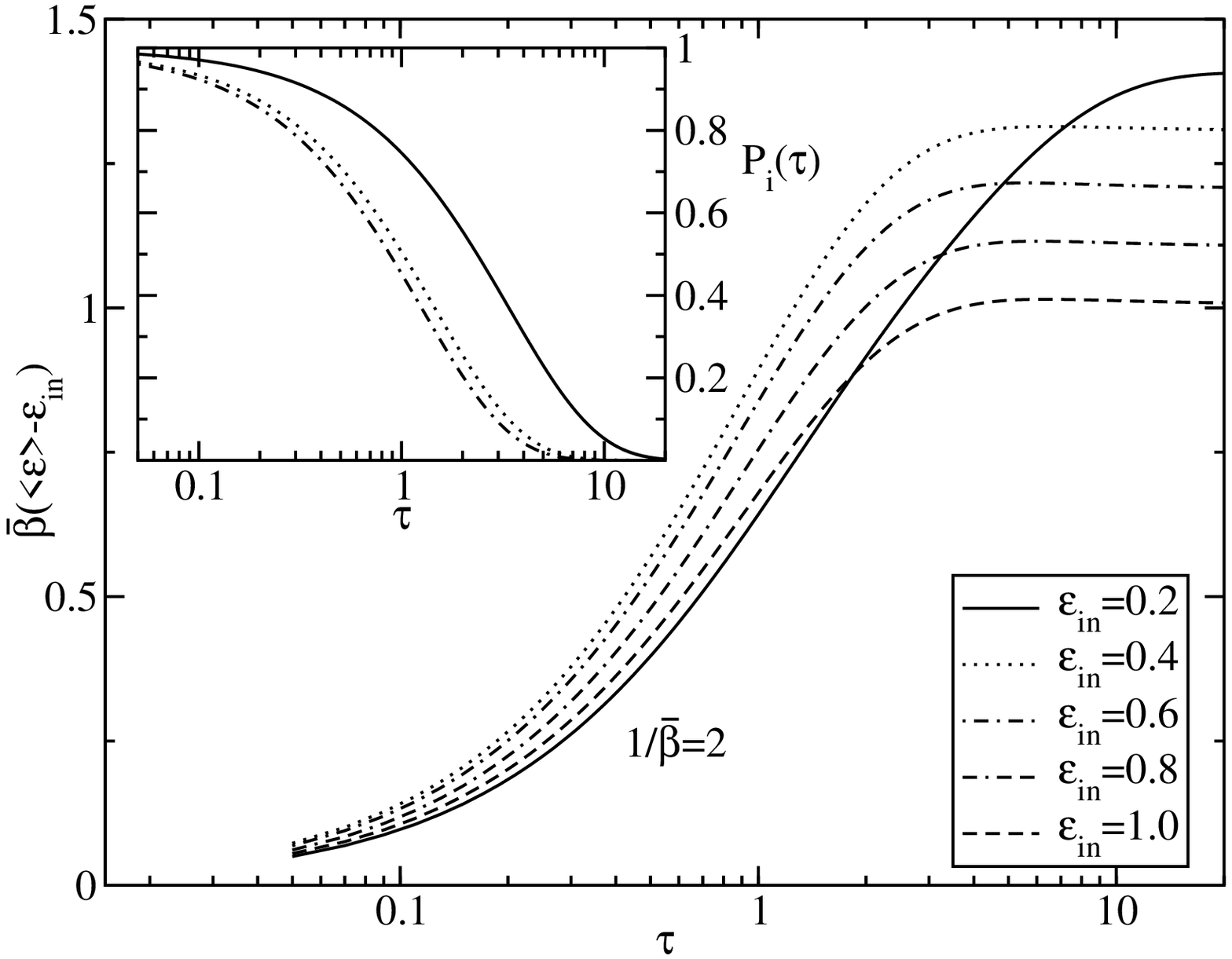,width=5.0cm,angle=-90}
\caption{Average energy acquired by the impurity in $k_B T$ unit as a function of $\tau$,
for some values of $\epsilon_{in}$.
In the inset is reported the relative
density $P_i(\tau)$ as a function of $\tau$. 
(a) $1/\bar\beta=10$, i.e. $k_BT=5\mu$. (b) $1/\bar\beta=2$, i.e. $k_BT=\mu$.
}
\label{fig1}
\end{figure}

In Fig.~\ref{fig1}a we report the average energy acquired by the impurity and lost
by the condensate thermal atoms,
in $k_BT$ unit, i.e. $\bar\beta[\langle\epsilon\rangle-\epsilon_{in}]=
\beta[\langle E\rangle-E_{in}]$. We have considered $1/\bar\beta=10$
and some values $(\ge1)$ of the initial impurity energy.
We can note that $\langle E\rangle=(3/2) k_B T$ for 
$\bar\tau\rightarrow\infty$, that is the average energy of a free particle at the 
equilibrium. In contrast, if we neglect the collective modes, a thermal particle of 
the condensate has an average energy 
equal to $\sim0.514 (3/2) k_BT$, that is, about one half of 
the classical value. It is easy to demonstrate that the number of lost thermal
atoms $\delta N$, when the condensate energy decreases by $\delta E$, is 
$\sim0.78\delta E/(k_BT)$. Therefore, if the impurity acquires an energy of 
$\sim1.3 k_BT$ a thermal atom jumps into the condensate state, balancing the loss
due to the impurity creation.
A nearly complete thermalization occurs for $\tau\simeq0.5$,
that is, for $t_{th}\simeq1/(2\sigma_0n_0 c)$, where $\sigma_0\equiv 4\pi a^2$. 
For a $^{23}Na$
condensate and a typical density of $10^{14}\div10^{15}cm^{-3}$ ($T\sim350\div3500nK$ for
$1/\bar\beta=10$) we have $t_{th}\simeq10\div0.3 ms$. 
\begin{figure}
\epsfig{figure=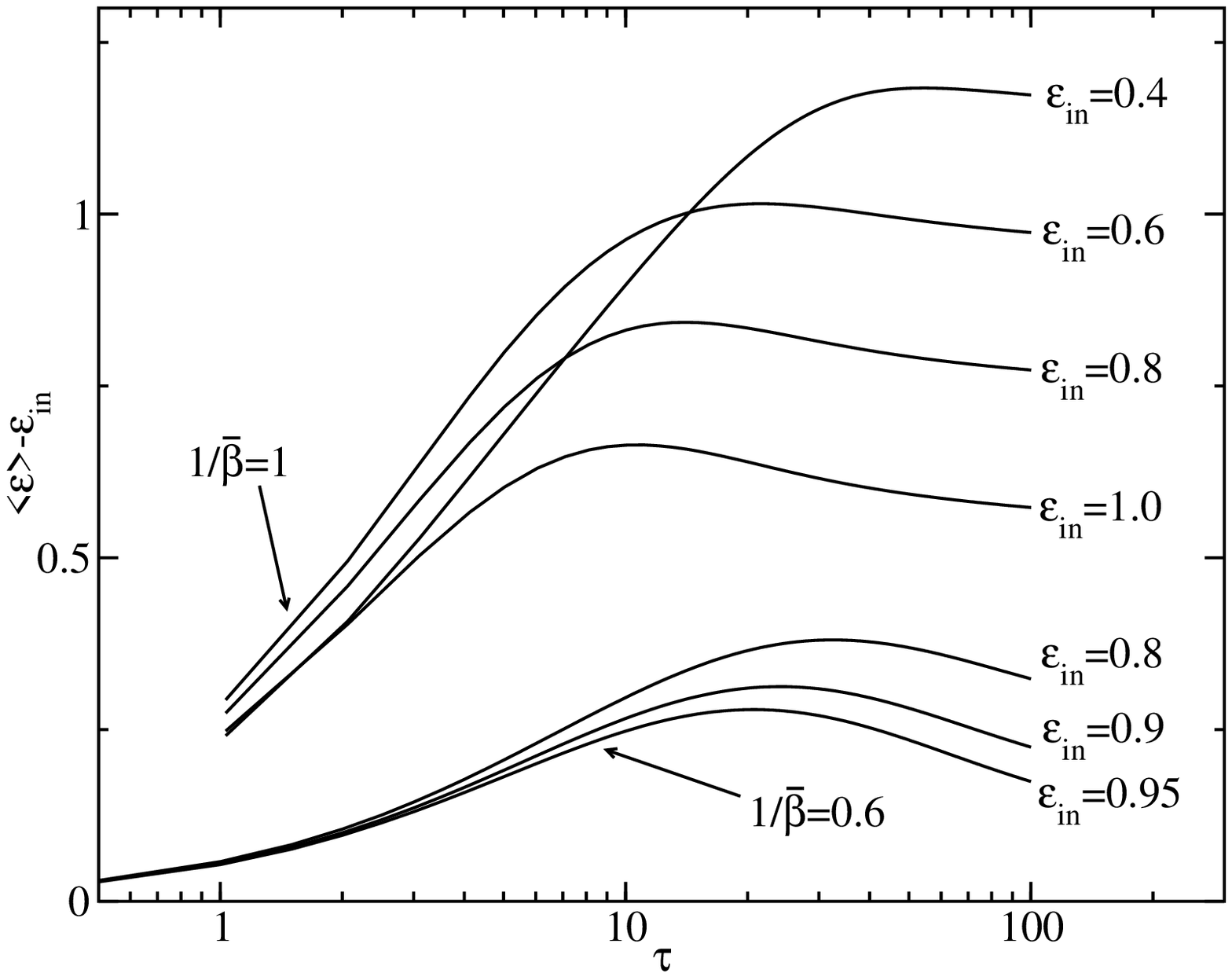,width=5.0cm,angle=-90}
\caption{Average energy acquired by the impurity in $\mu/2$ unit as a function of $\tau$,
for some values of $\epsilon_{in}$ and for $1/\bar\beta=0.6\div1$.
}
\label{fig2}
\end{figure}   
We can see that the thermalization time is independent on the initial velocity.
In fact, the impurity loses quickly the memory of its initial energy. In the inset
of Fig.~\ref{fig1}a we report the relative density 
$P_i(\tau)\equiv P(\epsilon_{in},\tau)/P(\epsilon_{in},0)$ as a function
of $\tau$. The initial state population decreases by $60\div70$ per cent for
$\tau\sim0.1$, i.e. for $t=0.06\div2 ms$. We note that the depletion rate is smaller
for a higher initial velocity, contrary to the zero temperature case. In fact, for $T=0$
we expect that the scattering rate grows upon increasing the initial velocity.
At finite temperatures, to reduce the laser mode depletion, it is convenient
to have a large initial velocity, because the scattering rate is smaller and, furthermore,
the impurity leaves the condensate earlier. For a sufficiently high velocity we expect
that the scattered impurity fraction is not enhanced by the finite temperature.

If the condensate temperature is reduced, the scattering rate decreases and approaches
the zero temperature value. For $T=0$ only the spontaneous scattering contributes 
and when $\epsilon_{in}<1$
the impurity thermalization is absent because
of the condensate superfluidity. 
For $k_BT=\mu$ ($\bar\beta=2$) the temperature can be still important. In fact,
we can see in Fig.~\ref{fig1}b that a nearly complete thermalization occurs for 
$\tau\sim2$, that, for $n_0=10^{15} cm^{-3}$, corresponds to $t\sim1.2ms$. The 
initial velocity is below the Landau value, so the first impurity scattering is 
due to the finite temperature. 
The function $G(\epsilon_f,\epsilon_i)$
is equal to zero for $\epsilon_f<1/\epsilon_i$, therefore, if the initial energy is
too low, the particle jumps in a state with a large energy. However 
the probability rate of this process is small, because of the presence of the exponential
in $G$. So we have a slowing down of the thermalization rate (see Fig.~\ref{fig1}b, 
solid line) when the initial velocity is very low. For $\epsilon_{in}\ge0.4$ the 
cut-off of the 
constraint $\epsilon_f>1/\epsilon_i$ has a minor effect and the thermalization time becomes
again independent on the initial velocity. Also the depletion rate of the relative
density $P_i(\tau)$ is reduced by the cut-off for $\epsilon_{in}=0.2$. Increasing
$\epsilon_{in}$ the depletion rate grows, however for $\epsilon_{in}>0.6$ it
slows down again, as it occurs for $\bar\beta=10$ (see Fig.~\ref{fig1}a).
For $\epsilon_{in}=0.4$ and $n_0=10^{14}\div10^{15}cm^{-3}$ the initial impurity
velocity is $\sim0.3\div1 cm/s$, therefore, the average distance before the
first collision ($\tau\sim1$) is about $4\cdot(10^{-2}\div10^{-3})mm$.  The radius of 
a condensate with the same central density is $\sim0.5\div2\cdot10^{-1}mm$ when the trap
frequency is $20Hz$.
To contrast the gravitational acceleration of the impurities, both the parent and
outcoupled condensates can be trapped in the radial direction with a far off-resonant
Gaussian laser beam. In the axial horizontal direction a magnetic field can be used to
trap {\it only} the parent condensate.
We can set the trap 
in a such way  to permit that the impurities leave the condensate after the 
thermalization time.

In Fig.~\ref{fig2} we report the same curves of Fig.~\ref{fig1}, but the energy
is in $\mu/2$ unit and $1/\bar\beta=0.6\div1$. In this case the energy reaches a
maximum at an intermediate time. For $1/\bar\beta=1$ and $\epsilon_{in}=0.6\div1.0$
the maximum is reached at $\tau=10\div20$, that, for $n_0=10^{15}cm^{-3}$, corresponds
to $t=6\div10ms$. We find that the evolution of the initial state 
population is nearly independent on $\epsilon_{in}$. At $\tau=10$ the initial state 
is depleted by $40\%$.
To cool the condensate it is not necessary to reach the maximum,
but for smaller times we obtain a minor reduction of the condensate temperature
per atom. Note that for $1/\bar\beta=0.6$ the equilibrium thermal energy is $0.9$,
in $\mu/2$ unit. Therefore, for $\epsilon_{in}=0.9\div0.95$ the initial energy
is equal or higher than the equilibrium value. Even so, at an intermediate time the
impurity acquires energy and the condensate cools. That occurs because when the velocity
is below the critical value the first scattering can only enhance the impurity energy.

Let us suppose now that there are many impurities, then many atoms can be
in the same energy level
and, therefore, it is necessary to consider the stimulation effects.
If we suppose that the initial impurity velocity is isotropic, we can make 
the following modification in Eq.~(\ref{bolteq})
\bey\nonumber
\frac{dp(\epsilon)}{d\tau}=[p(\epsilon)/\rho(\epsilon)+1]
\int G(\epsilon,\bar\epsilon)p(\bar\epsilon)d\bar\epsilon \\
\label{bolteq2}
-\int [p(\bar\epsilon)/\rho(\bar\epsilon)+1]G(\bar\epsilon,\epsilon)
p(\epsilon)d\bar\epsilon\equiv B(p),
\eey
where $p(\epsilon)=P(\epsilon)/V$, $V$ being the system volume, and
$\rho(2E/\mu)=(2m)^{3/2}\hbar^{-3}(2\pi)^{-2}\sqrt{E}$
is the level density per unit volume~\cite{huang,dalfovo}.
It is easy to demonstrate that the Bose-Einstein distribution 
$p(\epsilon)\propto\sqrt{\epsilon}(e^{\bar\beta(\epsilon-\bar\mu)}-1)^{-1}$ is a
steady solution of Eq.~(\ref{bolteq2}), $\bar\mu$ being the normalized chemical
potential of the impurity cloud.
\begin{figure}[t]
\epsfig{figure=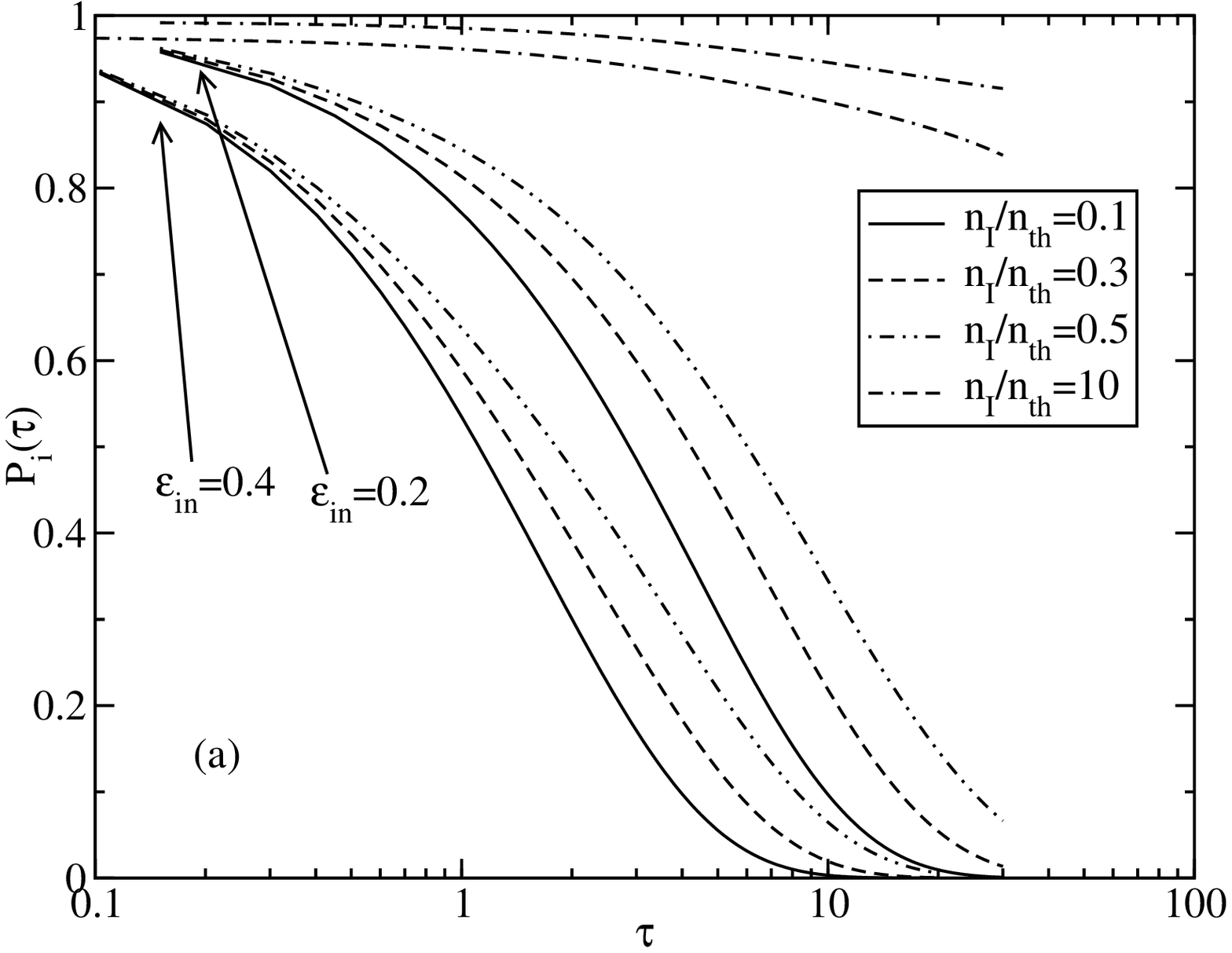,width=5.0cm,angle=-90}
\epsfig{figure=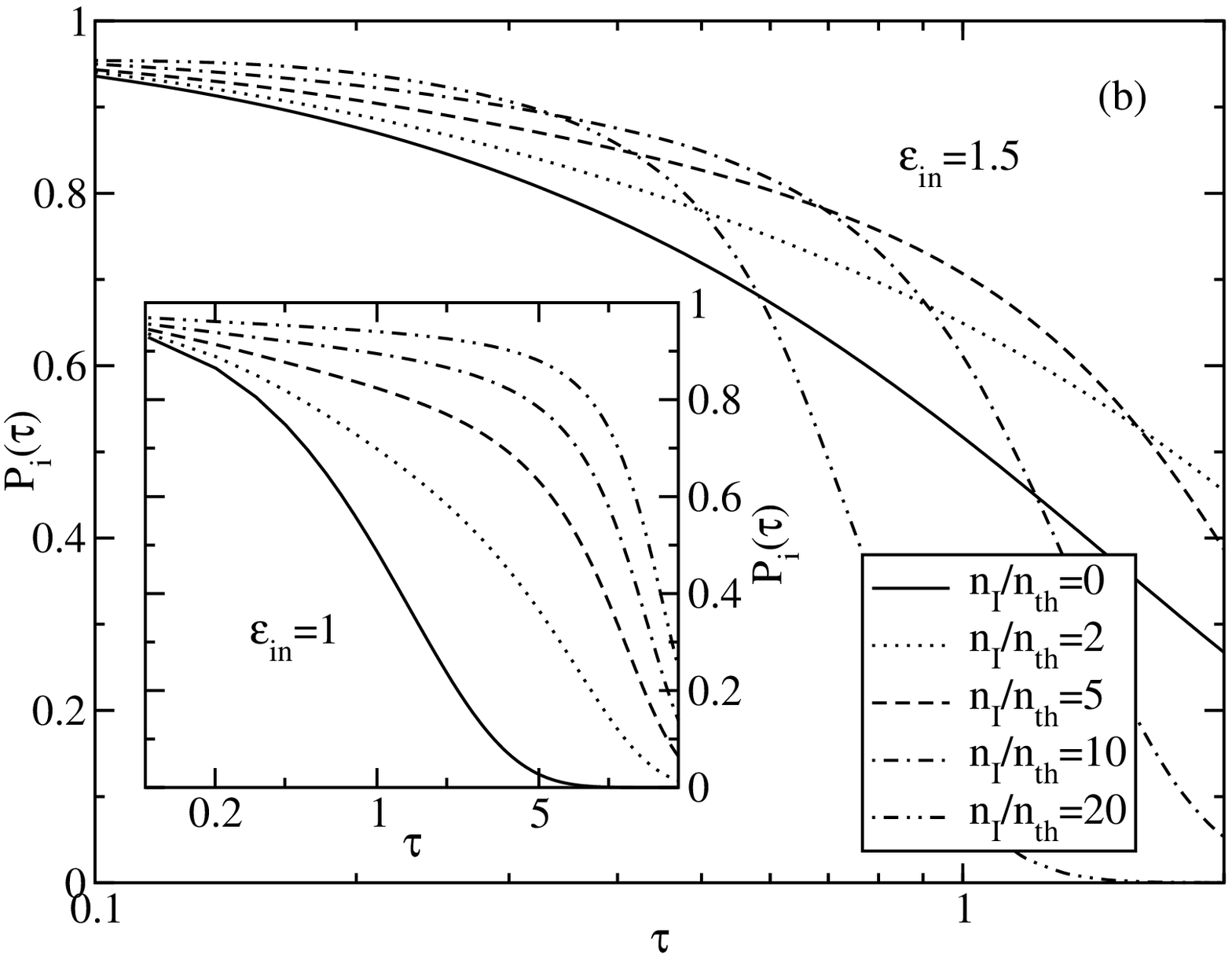,width=5.0cm,angle=-90}
\caption{Relative density $P_i(\tau)$ as a function of $\tau$ for $1/\bar\beta=2$ and 
some values of 
$n_I/n_{th}$, with (a) $\epsilon_{in}=0.2\div0.4$, (b) $\epsilon_{in}=1.5$ and
$\epsilon_{in}=1$ in the inset.}
\label{fig3}
\end{figure} 

We isolate in $p(\epsilon)$ the contribution of the initial populated state with
energy $\epsilon_{in}$. The associated distribution, that is very narrow, 
has a width equal to $\delta\epsilon\equiv2/(V\rho\mu)$. We indicate with $p_i$ its 
height.

Eq.~(\ref{bolteq2}) becomes 
\bey\label{bolteq3}
\nonumber
\frac{dp(\epsilon)}{d\tau}=B(p)+\delta\epsilon[p(\epsilon)/\rho(\epsilon)
+1]G(\epsilon,\epsilon_{in})p_i \\
-\delta\epsilon[p_i/\rho(\epsilon_{in})+1]G(\epsilon_{in},\epsilon)p(\epsilon) \\
\nonumber
\frac{dp_i}{d\tau}=[p_i/\rho(\epsilon_i)+1]
\int G(\epsilon_{in},\bar\epsilon)p(\bar\epsilon)d\bar\epsilon \\
\label{bolteq4}
-\int [p(\bar\epsilon)/\rho(\bar\epsilon)+1]G(\bar\epsilon,\epsilon_{in})
p_id\bar\epsilon.
\eey
The quantities $\delta\epsilon G(\epsilon_{in},\epsilon)p(\epsilon)$ and 
$\int G(\epsilon_{in},\bar\epsilon)p(\bar\epsilon)d\bar\epsilon$ 
in the first and second equation, respectively, can be neglected.
Therefore Eqs.~(\ref{bolteq3},\ref{bolteq4}) become
\bey\label{bolteq5}
\nonumber
\frac{dp(\epsilon)}{d\tau}=B(p)+\frac{2}{\mu}\{[p(\epsilon)/\rho(\epsilon)+1]
G(\epsilon,\epsilon_{in}) \\
-G(\epsilon_{in},\epsilon)p(\epsilon)/\rho(\epsilon_{in})\}n_i \\
\label{bolteq6}
\frac{dn_i}{d\tau}=-\gamma_i(p) n_i,
\eey
where $n_i$ is the spatial density of the impurity cloud and
\bey
\nonumber
\gamma_i(p)= 
\int [p(\bar\epsilon)/\rho(\bar\epsilon)+1]G(\bar\epsilon,\epsilon_{in})
d\bar\epsilon- \\
\nonumber
1/\rho(\epsilon_{in})\int g(\epsilon_{in},\bar\epsilon)p(\bar\epsilon)d\bar\epsilon=
\int\{G(\bar\epsilon,\epsilon_{in})- \\
W(\epsilon_{in},\bar\epsilon)\rho(\epsilon_{in})^{-1}[1-
(1+\Delta\epsilon^2/4)^{-1/2}]p(\bar\epsilon)\}d\bar\epsilon,
\eey
$\Delta\epsilon$ being $\epsilon_{in}-\bar\epsilon$. The first term of $\gamma_i$ is
the depletion rate of the initial state when the stimulated effects of the impurities
are negligible. The second term, that does not depend directly on the temperature, 
is proportional 
to the density $p(\bar\epsilon)$ and it is important when there are many impurities. 
This term is negative when 
$\epsilon_{in}<1$, therefore in this case it reduces the depletion rate $\gamma_i$.
In fact, the scattered atoms can have a high probability to scatter back into the original
macroscopically populated state.
On the contrary, when $\epsilon_{in}>1$ the second integrand is negative for 
$\bar\epsilon>\epsilon_{in}$ and positive for $\bar\epsilon<\epsilon_{in}$; if
$\epsilon_{in}$ is sufficiently high the impurity stimulation effect enhances the
depletion rate, as it has been experimentally observed~\cite{ketterle}.
In our calculations we have considered the condensate as a thermal reservoir, however
if we have many impurities the collisions can cool or warm the condensate, consequently
the depletion rate can decrease or increase. Our approximation is reasonable when
the density of outcoupled atoms is small with respect to the density of thermal atoms 
$n_{th}$ or, otherwise, for sufficiently small times, when the collided fraction
is small.
We have approximatively, if we neglect the condensate collective modes,~\cite{huang}
\be
n_{th}=(4\pi^2\hbar^3)^{-1}m^{3/2}(\mu/\bar\beta)^{3/2}
\ee
\begin{figure}[t]
\epsfig{figure=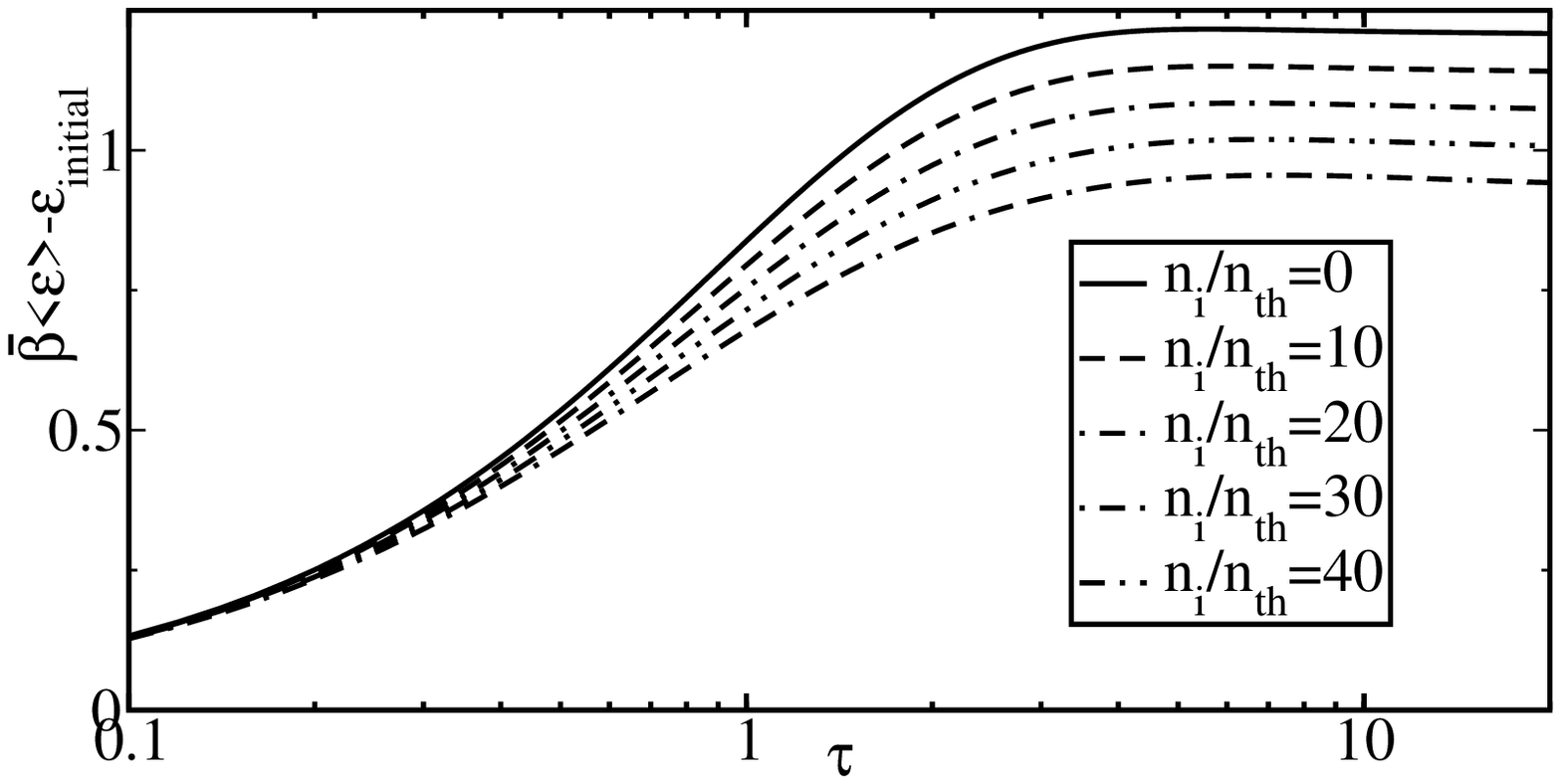,width=3.5cm,angle=-90}
\caption{Same as with Fig.~\ref{fig1}, but for an impurity cloud. Some values of $n_I$
are considered, for $1/\bar\beta=2$ and $\epsilon_{in}=0.6$.}
\label{fig4}
\end{figure} 
In Fig.~\ref{fig3}a we report the relative density $P_i(\tau)$ for $1/\bar\beta=2$, 
$\epsilon_{in}=0.2\div0.4$ and some values of $n_I/n_{th}\equiv n_i(0)/n_{th}$.
It is easy to demonstrate that these curves are independent on $\mu$ and, therefore,
on $a$ and $n_0$. In fact, with $p(\epsilon)\propto\sqrt{\mu}$ and 
$n_I\propto\mu^{3/2}$, the Boltzmann equations are invariant with respect to 
modifications of the chemical potential.
For $n_0=10^{14} cm^{-3}$ and $n_0=10^{15} cm^{-3}$ we have 
$n_{th}\simeq1.1\cdot10^{12} cm^{-3}$ and $n_{th}\simeq3.5\cdot10^{13} cm^{-3}$, 
respectively. 
We can see that for $n_I/n_{th}=10$ the
depletion rate is considerably reduced by the stimulation effects. In 
Fig.~\ref{fig3}b we plot the same curves but for $\epsilon_{in}=1$ (inset) and 
$1.5$, respectively. These curves are reliable only at the initial times because of
the cooling of the condensate. Furthermore, if collisions create many thermal 
impurity atoms, a lot of impurities
could condensate into the state with zero energy, but our equations are not suitable
to describe this phenomenon, because this state has to be treated separately from
the energy integral of Eq.~(\ref{bolteq5}). If the number of scattered 
atoms is lower than $n_{th}$ we can suppose that this condensation does not occur. 
Note that the depletion rate is lowered also for
$\epsilon_{in}=1.5$, that is above the Landau value. Instead, for 
$\epsilon_{in}=2$ we find that the depletion rate is enhanced. Increasing the temperature
the threshold of $\epsilon_{in}$ increases.

Finally, we report in Fig.~\ref{fig4} the same curves of Fig.~\ref{fig1}, but
accounting for the stimulation effects of the impurity cloud. We have considered
some values of $n_I$, for $1/\bar\beta=2$ and $\epsilon_{in}=0.6$. The maximum 
acquired energy is reduced increasing $n_I$. 

In conclusion, we have evaluated the thermalization time of impurity atoms inside
a condensate at finite temperature and we have found that their thermalization, that
occurs also below the Landau critical velocity, can cool the condensate. 
In fact, a {\it small} Raman or rf outcoupling creates a condensate in another level with
a lower temperature, that cools sympathetically the parent condensate.
We have shown that this effect
can be studied experimentally.
It could be used to remove the residual thermal atoms.
Furthermore, we have accounted for the stimulation effects of the impurity cloud and
we find that below the Landau critical velocity the macroscopic population of the
initial state can reduce drastically its depletion rate. For sufficiently high velocities
the opposite effect occurs, as observed in Ref.~\onlinecite{ketterle}. 
The threshold velocity depends on the condensate temperature.
Our analysis is also important for the development of intense atom lasers.

\end{document}